\documentclass[english,reqno,dvips,12pt]{article}
\usepackage[T2A]{fontenc}
\usepackage[cp1251]{inputenc}
\usepackage[english]{babel}
\usepackage{epsfig,graphicx,amsmath,amsfonts,verbatim,amssymb,amsthm}

\allowdisplaybreaks

\theoremstyle{plain}

\theoremstyle{remark}

\newcommand{\dt}{\delta}

\begin{document}

\title{Bose Condensate
in the $D$-Dimensional Case, in Particular, for $D=2$.
Semiclassical transition to the classical thermodynamics}
\author{V.~P.~Maslov}

\date{}

\maketitle

\begin{abstract}
The number-theoretical problem of partition of an integer corresponds to $D=2$.
This problem obeys the Bose--Eeinstein statistics, where repeated terms are admissible in the partition,
and to the Fermi--Dirac statistics, where they are inadmissible.
The Hougen--Watson P,Z-diagram shows that this problem splits into two cases:
the positive pressure domain corresponds to the Fermi system,
and the negative, to the Bose system. This analogy can be applied to the van der Waals thermodynamics.

The thermodynamic approach is related to four potentials corresponding to the energy, free energy, thermodynamic Gibbs potential, enthalpy.
The important notion of de Broglie's wavelength permits passing from particle to wave packet, in particular,
to Bose and Fermi distributions.

Particles of ideal Bose and Fermi gases
and the decay of a boson consisting of two fermions into separate fermions are studied.

The case of finitely many particles $N$ of the order of $10^2$ is considered
by heuristic considerations like those Fock used to derive
the Hartree--Fock equation.

The case of $N\ll1$ is studied by Gentile statistics, tropical geometry and
nonstandard analysis (Leibnitz differential or monad).

A relation for the energy of neutron separation from the atomic nucleus is obtained
when the atomic nucleus volume and de Broglie's wavelength are known.
The Appendix is author's paper written in 1995.
\end{abstract}

In 1925, Einstein, when examining a work of Bose, discovered a new phenomenon,
which he called the Bose condensate. A modern presentation of this discovery
can be found in~\cite{Landau_St-ph}. An essential point in this presentation is
to define the entropy of the Bose gas. The definition is related to the
dimension by means of the so-called ``number of states'' (cells), which is
denoted by $G_j$ in the book~\cite{Landau_St-ph}. After this, the problem of
minimizing the entropy is considered by using the Lagrange multipliers under
two constraints, namely, for the number of particles and for energy. The number
of states $G_j$ is determined by the formula which mathematicians call the
``Weyl relation;'' it is described in detail in~\cite{Landau_Quan-mech} in the
``semiclassical case'' in the section ``Several degrees of freedom.'' The
$2D$-dimensional phase space is partitioned into a lattice, and the number
$G_j$ is defined by the formula
\begin{equation}\label{l1}
G_i=\frac{\Delta p_j \Delta q_j}{(2\pi h)^D}.
\end{equation}
The indeterminate Lagrange multipliers are expressed in terms of temperature
and chemical potential of the gas.

Further, in~\cite{Landau_St-ph}, following Einstein, a passage to the limit is
carried out as $N\to\infty$, which enables one to pass from sums to integrals.
Then, in the section ``Degenerate Bose gas,'' a point is distinguished which
corresponds to the energy equal to zero. This very point is the point of Bose
condensate on which excessive particles whose number exceeds some value
$N_d\gg1$ are accumulated at temperatures below the so-called degeneracy
temperature $T_d$. The theoretical discovery of this point anticipated a number
of experiments that confirmed this fact not only for liquid helium but also for
a series of metals and even for hydrogen.

From a mathematical point of view, distinguishing a point in the integral is an
incorrect operation if this point does not form a $\delta$ function. In
particular, for the two-dimensional case, this incorrectness leads to a
``theorem'' formulated in various textbooks and claiming that there is no Bose
condensate in the two-dimensional case.

In this paper, we get rid of this mathematical incorrectness and show that,
both in the two-dimensional and in the one-dimensional case, the Bose
condensate exists if the point introduced above is well defined.

If we accept Einstein's remarkable discovery for the three-dimensional case and
justify it in a mathematically correct way, then the Bose condensate in the
two-dimensional case is equally correct mathematically. We dwell on the
two-dimensional case below in particular detail.

 In physics, the Bose--Einstein and Fermi--Dirac distributions are determined
 by using the Gentile statistics~\cite{Gentile} (parastatistics).
The Gentile statistics comprises the Bose statistics and the Fermi statistics as special cases.
The Gentile statistics contains an additional constant~$k$ which denotes the maximal number of particles
located at a fixed energy level. In particular, for $k=1$, the distributions of the Gentile statistics
coincide with the distributions of the Fermi--Dirac statistics.
In the Gentile statistics, one assumes that $k\geq1$.

Thus, we  first consider the case in which $N\gg1$, but $n$ is not equal to infinity.
In the section ``Ideal gas in the case of parastatistics'' of the textbook by
Kvasnikov~\cite{Kvasn}, there is a problem (whose number in the book is~(33))
which corresponds to the final parastatistics
\begin{equation}\label{kvas}
n_j= \frac{1}{\exp\{\frac{\varepsilon_j-\mu}{T}\} -1} -
\frac{k+1}{\exp\{(k+1)\frac{\varepsilon_j-\mu}{T}\} -1}, \qquad
n_j=\frac{N_j}{G_j}.
\end{equation}
In our case, we have $k=N_d$, and the point of condensate is $\varepsilon_0=0$.

By~\eqref{l1}, it is clear that $G_j$ is associated with the $D$-dimensional
Lebesgue measure and, in the limit with respect to the coordinates $\Delta
q_j$, gives the volume~$V$ in the space of dimension~$3$ and the area $Q$ in
the space of dimension~$2$. The passage with respect to the momenta $\Delta
p_j$ is also valid as $N\to\infty$ and $\mu>\delta>0$, where $\delta$ is
arbitrarily small.

Expanding~\eqref{kvas} at the point $\varepsilon_0=0$ in the small parameter
$$x=(\mu N_d)/T_d,$$ where $N_d$ stands for the number of particles
corresponding to the degeneration and $T_d$ for the degeneracy temperature, and
writing $$\xi=-\mu/T_d,
$$
we obtain $G_0=1$, see~\eqref{3.9} below,
\begin{gather}\label{dist1} n_0= \bigg\{  \frac{1}{\exp\{\frac{-\mu}{T}\} -1}
- \frac{N_d+1}{\exp\{(N_d+1)\frac{-\mu}{T}\} -1} \bigg\}=
\frac{e^{\xi N_d}-1 -(N_d+1)(e^{\xi}-1)}{(e^{\xi}-1)(e^{2N_d}-1)}\notag\\
=\frac{N_d}{2}\frac{1+\frac{x}{6}+\frac{x^2}{4!}+\frac{x^3}{5!}+ \dots}
{1+\frac{x}{2}+\frac{x^2}{6}+\frac{x^3}{4!}+ \dots}=
\frac{N_d}{2}\bigg(1-\frac{x}{3}-\frac{11}{24} x^2  -0.191 x^3-\dots\bigg).
\end{gather}

For example, if $x\to0$, then $n_0=N_d/2$, and hence the number $n_0$ in
the condensate at $T=T_d$ does not exceed $N_d/2$. If $x=1.57$, then
$n_0\approx N_d/10$. Certainly, this affects the degeneracy temperature,
because this temperature can be expressed only in terms of the number of
particles above the condensate, $\tilde{N}_d$, rather than in terms of the
total number of particles $N_d$ (which is equal to the sum of $\tilde{N}_d$ and
of the number of particles in the condensate).

According to the concept of Einstein, at $T=T_d$ the condensate contains
$o(N_d)$ particles. However, even this accumulation gives a $\delta$ function,
albeit with a small coefficient, for example, $\tilde{N}_d/\ln N_d$.

To reconcile the notion of Bose statistics which is given
in~\cite{Landau_St-ph} with symmetric solutions of the $N$-particle
Schr\"odinger equation, i.e., of the direct sum of $N$ noninteracting
Hamiltonians corresponding to the Schr\"odinger equation, and the symmetric
solutions of their spectrum, it is more appropriate to assign to the cells the
multiplicities of the spectrum of the Schr\"odinger equation in the way
described in~\cite{FAN-2003}.

Consider the nonrelativistic case in which the Hamiltonian $H$ is equal to
$${p^2}/({2m}),$$ where $p$ stands for the momentum.

The comparison of $G_i$ with the multiplicities of the spectrum of the
Schr\"odinger equation gives a correspondence between the eigenfunctions of the
$N$-partial Schr\"odinger equation that are symmetric with respect to the
permutations of particles and the combinatorial calculations of the Bose
statistics that are presented in~\cite{Landau_St-ph}.

A single-particle $\psi$-function satisfies the free Schrodinger equation with
the Dirichlet conditions on the vessel walls. According to the classical
Courant formula,
 \begin{equation}\label{cour-1}
\lambda_j\sim \frac{2h^2}{m}\bigg(\frac{\pi^{D/2}\Gamma(D/2+1)}{V}\bigg)^{2/D}
j^{2/D} \qquad \text{as} \quad j\to\infty,
\end{equation}
where $D$ stands for the dimension of the space, because the spectral density
has the asymptotic behavior
\begin{equation}\label{cour-2}
\rho(\lambda)= \frac{Vm^{D/2}\lambda^{D/2}}{\Gamma(D/2+1)(2\pi)^{D/2} h^D}
(1+o(1)) \qquad \text{as} \quad \lambda\to\infty.
\end{equation}
The asymptotics~\eqref{cour-1} is a natural generalization of this formula.

Using this very correspondence, we establish a relationship between the
Bose--Einstein combinatorics~\cite{Landau_St-ph}, the definition of the
$N$-particle Schr\"odinger equation, and the multiplicity of the spectrum of
the single-particle Schr\"odinger equation.

The spectrum of the single-particle Schr\"odinger equation, provided that the
interaction potential is not taken into account, coincides, up to a factor,
with the spectrum of the Laplace operator. Consider its spectrum for the closed
interval, for the square, and for the $D$-dimensional cube with zero boundary
conditions. This spectrum obviously consists of the sum of one-dimensional
spectra.

On the line we mark the points $i=0,1,2,\dots$ and on the coordinate axes $x,y$
of the plane we mark the points with $x=i=0,1,2,\dots$ and $y=j=0,1,2,\dots$.
To this set of points $(i,j)$ we assign the points on the line that are
positive integers, $l=1,2,\dots$.

To every point we assign a pair of points, $i$ and $j$, by the rule $i+j=l$.
The number of these points is $n_l=l+1$. This is the two-dimensional case.

Consider the 3-dimensional case. On the axis $z$ we set $k=0,1,2,\dots$, i.e.,
let $$i+j+k=l$$ In this case, the number of points $n_l$ is equal to
$$n_l=\frac{(l +1)(l +2)}{2}.$$

It can readily be seen, for the $D$-dimensional case, that the sequence of
multiplicities for the number of variants $$i=\sum_{k=1}^Dm_k,$$ where $m_k$
are arbitrary positive integers, is of the form
\begin{equation}\label{3.7}
q_i(D) = \frac{(i+D-2)!}{(i-1)!(D-1)!}, \qquad \text{for} \ D=2, \quad
q_i(2)=i,
\end{equation}
\begin{equation}\label{3.8}
\sum_{i=1}^\infty N_i=N, \qquad  \varepsilon \sum_{i=1}^\infty  q_i(D) N_i=E.
\end{equation}

The following problem in number theory corresponds to the three-dimensional
case $D=3$ (cf.~\cite{Landau_St-ph}):
\begin{equation}\label{dist_4}
\sum_{i=1}^\infty  N_i =N, \qquad  \varepsilon\sum_{i=1}^\infty
\frac{(i+2)!}{i!6} N_i = E, \qquad  \frac{E}{\varepsilon}=M.
\end{equation}

Write $M=E_d/\varepsilon_1$, where $\varepsilon_1$ stands for the coefficient
in formula~\eqref{cour-1} for $j=1$. Let us find $E_d$,
\begin{equation}\label{cour-3}
E_d= \int_0^\infty \frac{\frac{|p|^2}{2m}\,
d\varepsilon}{e^{\frac{|p|^2}{2m}/T_d}-1},
\end{equation}
where
\begin{equation}\label{cour-4}
d\varepsilon=\frac{|p|^2}{2m} \frac{dp_1 \dots dp_D \, dV_D}{(2\pi h)^D}.
\end{equation}
Whence we obtain the coefficient $\alpha$ in the formula,
\begin{equation}\label{cour-5}
E_d =\alpha T_d^{2+\gamma}\zeta(1+D/2)\gamma(1+D/2),
\end{equation}
where $\gamma=D/2-1$.

To begin the summation in~\eqref{3.8} at the zero index (beginning with the
zero energy), it is necessary to rewrite the sums~\eqref{3.8} in the form
\begin{equation}\label{3.9}
\sum_{i=0}^\infty N_i=N, \qquad  \varepsilon \sum_{i=0}^\infty  (q_i(D)-1)
N_i=E-\varepsilon N.
\end{equation}

The relationship between the degeneracy temperature and the number
$\tilde{N}_d$ of particles above the condensate for $\mu>\delta>0$ (where
$\delta$ is arbitrarily small) can be found for $D>2 $ in the standard way.

Thus, we have established a relationship between $G_i$ in formula~\eqref{l1}
(which is combinatorially statistical) and the multiplicity of the spectrum for
the single-particle Schr\"odinger equation, i.e., between the
statistical~\cite{Landau_St-ph} and quantum-mechanical definitions of Bose
particles.

For $D = 2$, the general problem reduces to a number theory problem.

Consider the two-dimensional case in more detail. There is an Erd\H os' theorem
for a system of two Diophantine equations,
\begin{equation}\label{df-eq}
\sum_{i=1}^\infty N_i=N, \qquad  \sum_{i=1}^\infty i N_i=M.
\end{equation}
 The maximum number of solutions of this system is achieved
if the following relation is satisfied:
\begin{equation}\label{8}
N_d=c^{-1}M_d^{1/2} \log\, M_d+ aM_d^{1/2}+o(M_d^{1/2}), \qquad
c=\pi\sqrt{2/3},
\end{equation}
and if the coefficient $a$ is defined by the formula $$c/2=e^{-ca/2}.$$

The decomposition of $M_d$ into one summand gives only one version. The
decomposition $M_d$ into $M_d$ summands also provides only one version (namely,
the sum of ones). Therefore, somewhere in the interval must be at least one
maximum of the variants. Erd\H os had evaluated it~\eqref{8}
(see~\cite{Erdos-46}).

If the number $N$ increases and $M$ is preserved in the problem~\eqref{df-eq},
then the number of solutions decreases. If the
sums~\eqref{df-eq} are counted from zero rather than from one, i.e., if we set
\begin{equation}\label{6} \sum_{i=0}^\infty  i N_i = (M-N), \qquad
\sum_{i=0}^\infty  {N_i} = N,
\end{equation}
then the number of solutions does not decrease and remains constant.

We shall  try to explain this effect. The
Erd\H os--Lehner problem~\cite{Erdos-Leh} is to decompose $M_d$ into $N\leq
N_d$ summands. Let us expand the number $5$ into two summands. We obtain
$3+2=4+1$. The total number is 2 versions (this problem is known as ``partitio
numerorum''). If we include 0 to the possible summands, we obtain three
versions: $5+0=3+2=4+1$. Thus, the inclusion of zero makes it possible to say
that we expand a number into $k\leq n$ (positive integer) summands. Indeed, the
expansion of the number $5$ into three summands includes all the previous
versions, namely, $5+0+0$, $3+2+0$, and $4+1+0$, and adds new versions, which
do not include zero.

In this case, the maximum number of versions for the
decomposition of the number $5$ into $N$ summands (there are two versions) is
achieved at $N=2$ and $N=3$ (the two values for the maximum number of versions
for $N$ above the condensate).

In this case, the maximum does not change drastically~\cite{Erdos-Leh};
however, the number of versions is not changed, namely, the zeros, i.e., the
Bose condensate, make it possible that the maximum remains constant, and the
entropy never decreases; after reaching the maximum, it becomes constant. This
remarkable property of the entropy enables us to construct an unrestricted
probability theory in the general case~\cite{MN_91-5}.

 Let us write the thermodynamic potential
of the system of noninteracting particles
in discrete form~\cite{Landau_St-ph}
\begin{equation}\label{v1:v570}
\Omega=\frac{-T}{\Lambda^{2(1+\gamma)}}
\sum_i \ln \sum_{n=0}^N g_i\left(\exp \frac{\mu-ih\omega}{T}\right)^n,
\end{equation}
where
$T$
is the temperature,
$i$
is the energy
in the
$i$th state
in the oscillatory ``representation,''
$\mu$
is the chemical potential,
$g_i$
is
the statistical weight
of the
$i$th state,
$h$
is the Planck constant,
$\omega$
is the oscillation frequency,
and
$\Lambda$
is the dimensionless quantity
depending on the particle mass~\cite{MN_93-1:v570}.

 Let
\begin{equation}\label{v2:v570}
\Omega_i =\frac{- T}{\Lambda^{2(1+\gamma)}}\ln \sum_{n=0}^N g_i \left(\exp \biggl(\frac{\mu-ih\omega}{T}\biggr)\right)^n, \qquad g_i= i^{D/2}.
\end{equation}
 For
Bose systems,
it is usual to assume
that
$N_i$
can take
any integer values
$0,1, 2, \dots$
up
to infinity.
 Next,
the infinite geometric sequence is summed, etc.
 The sum over n in formula~\eqref{v2:v570} is taken up to $N$,
which is the total number of particles
in the system.

 Let us construct a thermodynamics of the type of an ideal Bose gas
with a bounded number of states
at a given quantum level.
 The left-hand equality in the formula
\begin{equation}
\label{ll-1:v570}
\sum_j N_j =\sum_j G_j \bar{n}_j =N, \qquad
\sum_j \varepsilon_jN_j =\sum_j \varepsilon_j G_j \bar{n}_j =E,
\end{equation}
implies that
$N_i \leq N$;
hence
this condition
is not an additional constraint.
 Summing
the finite geometric
sequence, we obtain
\begin{equation}\label{v3:v570}
\Omega_i(k)= \frac{-T}{\Lambda^{2(1+\gamma)}}
\ln\sum_{n=0}^N g_i\left(\exp \biggl(\frac{\mu -ih\omega}{T}\biggr)\right)^n
=\frac{1}{\Lambda^{2(1+\gamma)}}\ln g_i
\left(\frac{1-\exp \bigl(\frac{\mu -ih\omega}{T}\bigr)(N+1)}
{1-\exp\bigl(\frac{\mu -ih\omega}{T}\bigr)}
\right).
\end{equation}

 The potential~$\Omega$
is equal to the sum of the~$\Omega_i$
over~$i$:
\begin{equation}
\label{v4:v570}
\Omega= \sum \Omega_i,
\qquad d\Omega=-S\, dT-N\, d\mu.
\end{equation}

 The number of particles
is
$N=-\partial \Omega/\partial\mu$.
 Hence we obtain
\begin{equation}
\label{v5:v570}
 N=\frac{1}{\Lambda^{2(1+\gamma)}}\sum_i\left(\frac{i^\gamma}
{\exp{\bigl(\frac{\mu -ih\omega}{T}}\bigr)-1}-
\frac{(N+1) i^\gamma}{\exp{\bigl((N+1)
\bigl(\frac{\mu -ih\omega}{T}\bigr)\bigr)}-1}\right).
\end{equation}

 The volume~$V$
was only needed for the purpose of normalization
in the passage from the number~$N$
of particles to their density.
 For
$\gamma>0$,
the volume~$V$
does
not affect the asymptotics as
$N\to\infty$,
because the term
containing the quantity
$N+1$
on the right-hand side
is small.

 But even,
for
$\gamma=0$,
in
view of Example~1 from~\cite{MN_93-1:v570},
a term of the form
$\ln N$
arises;
this term
must be taken
into account,
because,
in the two-dimensional case,
$\ln N \approx 15$.

 In the example referred to above,
we have
$D=2$,
$\gamma=0$,
but
there is no area~$\mathfrak{S}$.
 This feature is confusing for specialists
in thermodynamics.
 Indeed,
on the one hand,
$N/\mathfrak{S} \to \text{const}$,
but,
on the other hand,
it follows from
the same example
that
$\ln M_c \sim 2 \ln N_c$
and, therefore, the limit of
$N/\mathfrak{S}$
as
$N_c \to \infty$,
$\mathfrak{S}\to \infty$
tends
to infinity.
 This
ultimately leads
to the erroneous conclusion that,
in the two-dimensional case,
the Bose condensate exists
only
at
$T=0$.
 In fact,
it exists
at
$$
\text{
$T_d=\frac{h^2}{\sqrt{2}m}
\biggl(\frac{N}{\mathfrak{S}}\biggr)\frac{1}{\ln N}\mspace{2mu}$.
}
$$
(see below).

 In a two-dimensional trap, the number~$N$
is significantly smaller,
but, even
for
$N=100$,
$\ln N=2$,
we can apply asymptotic formulas given below.

 On the other hand, using relations
between the thermodynamic parameters,
we can decrease the number of independent variables
from three to two.

 Taking
the parameter~$\gamma$
into account,
we use the Euler--Maclaurin formula,
obtaining
$$
\sum_{j}\biggl(\frac{j^\gamma}{e^{bj+\varkappa}-1}
-\frac{kj^\gamma}{e^{bkj+k\varkappa}}\biggr) =\frac1{\alpha}\int^\infty_0
\biggl(\frac{1}{e^{bx+\varkappa}-1}-\frac{k}{e^{bkx+k\varkappa}-1}\biggr)\,dx^
\alpha+R,
$$
where
$\alpha=\gamma+1$,
$k=N+1$,
$b=1/T$,
and
$\varkappa= -\mu/T$.
 Here the remainder~$R$
satisfies the estimate
$$
|R|\leq \frac1\alpha\int^\infty_0|f'(x)|\,dx^\alpha, \qquad
\text{where}\quad
f(x)=\frac1{e^{bx+\varkappa}-1}-\frac{k}{e^{k(bx+\varkappa)}-1}\mspace{2mu}.
$$
 Calculating the derivative,
we obtain
\begin{align}
\label{ad1:v570}
f'(x)&=\frac{bk^2e^{k(bx+\varkappa)}}{(e^{k(bx+\varkappa)}-1)^2}
-\frac{be^{bx+\varkappa}}{(e^{bx+\varkappa}-1)^2}\mspace{2mu},
\nonumber\\[-3\jot]
\\
|R|&\leq \frac{1}{\alpha b^\alpha}
\int^\infty_0
\Big|\frac{k^2e^{k(y+\varkappa)}}{(e^{k(y+\varkappa)}-1)^2}
-\frac{e^{y+\varkappa}}{(e^{y+\varkappa}-1)^2}\Big|\,dy^\alpha.
\nonumber
\end{align}
 We also have
$$
\frac{e^z}{(e^z-1)^2}=\frac1{z^2}+\psi(z),
$$
where
$\psi(z)$
is a smooth function
and
$$
\text{
$|\psi(z)|\leq C(1+|z|)^{-2}$.
}
$$
 Setting
$z=y$
and
$z= ky$,
we obtain the estimate
\begin{align}
|R|&\leq \frac{1}{\alpha b^\alpha} \int^\infty_0
\big|\psi\big(k(y+\varkappa)\big)-\psi(y+\varkappa)\big|\,dy^\alpha
\nonumber\\&
\leq
\frac{k^{-\alpha}}{b^\alpha} \int^\infty_{k\varkappa} |\psi(y)|\,dy^\alpha
+\frac{1}{b^\alpha}\int^\infty_{\varkappa}|\psi(y)|\,dy \leq
 C b^{-\alpha},
\label{ad1a:v570}
\end{align}
where
$C$
is a constant.
 For example,
if
$\varkappa\sim(\ln k)^{-1/4}$,
then
$|R|$
preserves the estimate
$|R| \sim O(b^{-\alpha})$.

 The energy is denoted by~$M$,
because,
without multiplication by the volume~$V$,
we deal with a rather unusual thermodynamics,
which is, really, an analog of number theory
(see Example~1
from~\cite{MN_93-1:v570}).

 In evaluating~$M$,
we can neglect the correction in~\eqref{v3:v570},
and hence we obtain
\begin{equation}
\label{v8:v570}
 M=\frac {\Lambda^{\gamma_c-\gamma}}{\alpha\Gamma(\gamma+2)}
\int\frac{\xi\,d\xi^\alpha}{e^{b\xi}-1}
=\frac {\Lambda^{\gamma_c-\gamma}}{b^{1+\alpha}}
\int_0^\infty\frac{\eta\, d\eta^\alpha}{e^\eta-1},
\end{equation}
where
$\alpha=\gamma+1$,
$b=1/T_r$.
 Therefore,
$$
b=\frac1{M^{1/(1+\alpha)}}\left(\frac {\Lambda^{\gamma_c-\gamma}}{\alpha\Gamma(\gamma+2)}\int_0^\infty\frac{\xi
\,d\xi^\alpha}{e^\xi-1}\right)^{1/(1+\alpha)}.
$$

 We have
(see~\cite{Arxiv_2009:v570})
$$
\begin{aligned}
&\sum_j\biggl(\frac{j^\gamma}{e^{bj+\varkappa}-1}
-\frac{kj^\gamma}{e^{bkj+k\varkappa}}\biggr) =
\frac{1}{\alpha}\int_0^\infty
\left(\frac1{e^{b\xi}-1}-\frac{k}{e^{kb\xi}-1}
\right)\,d\xi^\alpha + O(b^{-\alpha}) \notag
\\&\qquad
=\frac{1}{\alpha b^\alpha}\int_0^\infty\left(\frac1{e^\xi-1}
-\frac1\xi\right)\,d\xi^\alpha
+\frac1{\alpha b^\alpha}\int_0^\infty\left(\frac1\xi-
\frac1{\xi(1+(k/2)\xi)}\right)\,d\xi^\alpha \notag
\\&\qquad\qquad
-\frac{k^{1-\alpha}}{\alpha b^\alpha}\int_0^\infty\left(
\frac{k^\alpha}{e^{k \xi}-1}
-\frac{k ^\alpha}{k \xi(1+(k /2)\xi)}\right)\,d\xi^\alpha
+ O(b^{-\alpha})\notag
\\ &\qquad
=\quad\frac{c(\gamma)}{b^\alpha}(k^{1-\alpha}-1)+O(b^{-\alpha}).
\end{aligned}
$$
 Setting
$k=N|_{\tilde{\mu}/T=0}\gg 1$,
we finally obtain
\begin{equation}
\label{ad2:v570}
 N|_{\tilde{\mu}/T=0}\cong (\Lambda^{\gamma_c-\gamma} c(\gamma))^{1/(1+\gamma)}T,
\qquad\text{where}\quad
c(\gamma)=\int_0^\infty\biggl(\frac1{\xi}-\frac1{e^\xi-1}\biggr)
\xi^\gamma\,d\xi.
\end{equation}

\bigskip

\textbf{The two-dimensional Bose condensate.}
 It can be proved that
$\varkappa\to 0$
gives the number
$N$
with satisfactory accuracy.
 Hence
$$
N_c=\int_0^\infty\bigg(\frac1{e^{bx}-1}
-\frac{N_c}{e^{bN_cx}-1}\bigg)\,dx +O(b^{-1}).
$$

 Consider the value of the integral (with the same integrand)
taken from
$\varepsilon$
to
$\infty$
and then pass to the limit as
$\varepsilon\to0$.
 After making the change
$bx=\xi$
in the first term and
$bN_cx=\xi$
in the second term, we
obtain
\begin{align}
 N_c&=\frac1b\int_{\varepsilon b}^\infty\frac{\,d\xi}{e^\xi-1}
-\int^\infty_{\varepsilon bN_c}\frac{\,d\xi}{e^\xi-1} +O(b^{-1})=\frac1b
\int^{\varepsilon bN_c}_{\varepsilon b}\frac{\,d\xi}{e^\xi-1}+O(b^{-1})\\&\sim \frac1b\int^{\varepsilon bN_c}_{\varepsilon
b}\frac{\,d\xi}\xi+O(b^{-1}) =\frac1b(\ln(\varepsilon
bN_c)-\ln(\varepsilon b))+O(b^{-1})=\frac 1b\ln N_c+O(b^{-1}). \label{ad3:v570}
\end{align}

 On the other hand, making the change
$bx=\xi$
in~\eqref{v8:v570}, we find that
$$
\frac1{b^2}\int^\infty_0\frac{\xi \,d\xi}{e^\xi-1}\cong M.
$$
 This yields
\begin{equation}
b=\bigg({\sqrt M}\bigg/{\sqrt{\int_0^\infty\frac{\xi
\,d\xi}{e^\xi-1}}}\,\bigg)^{-1}, \qquad N_c=\frac12\frac{\sqrt
 M}{\sqrt{\pi^2/6}}\ln M(1+o(1)) +O(b^{-1}).\label{ad4:v570}
\end{equation}

 Now let us find the next term of the asymptotics by setting
$$
N_c=c^{-1}M^{1/2}\ln c^{-1}M^{1/2}+O(b^{-1}) = c^{-1}M^{1/2}\ln N_c + O(b^{-1}),
\qquad\text{where}\quad c=\frac{2\pi}{\sqrt6}\,.
$$
 Passing to dimensionless quantities,
we obtain
\begin{equation}
 T_d\cong \frac{h^2}{\sqrt{2}m}
\frac{N_c}{\mathfrak{S}}\frac{1}{\ln N_c}\mspace{2mu},
\label{ad5:v570}
\end{equation}
where
$T_d$
is the degeneration temperature.

 Since physicists assume that
$N=\infty$,
they infer that
there is no Bose condensate for noninteracting particles
in the two-dimensional case.
 At the same time, since it was assumed that the superfluidity
of a weakly nonideal Bose gas is related to
the Bose condensate of an ideal Bose gas,
physicists infer that a sort of quasi-Bose-condensate arises.

\bigskip

\textbf{The case $N\ll 1$.}

Further, we consider the case where the number of gas molecule is $N\ll 1$.
We consider the neutrons and protons (nucleons) comprising the atomic nucleus of a molecule
from the thermodynamic point of view.
In particular, we use the de Broglie thermal wavelength
which determines the value of the wave packet corresponding to a given quantum particle.

The Hartree--Fock equation corresponding to weak interaction near the intersection of wave packets
allows one to write a self-consistent relation for the potential
which holds the nucleons in the nucleus and prevents the nucleus from decay.
Our goal is to apply thermodynamic methods related to the de Broglie thermal wavelength
and mathematical methods of the number theory and nonstandard analysis
to calculate the energy of neutron separation from the atomic nucleus.

By the concept of wave--particle duality, the corpuscule or wave character of a particle can be determined
by using a qualitative parameter, i.e., the de Broglie thermal wavelength.
If the de Broglie thermal wavelength is comparatively large, then the particle is a wave packet,
i.e., it is a quantum particle. In particular, such quantum particles in atomic physics
are called bosons or fermions.

If the de Broglie thermal wavelength and the nucleus volume are known,
one can determine the energy required for one neutron to separate from the nucleus
and for the nucleus to turn into a Fermi particle from a Bose particle
or, conversely,
from a Fermi particle into a Bose particle.
The energy is usually calculated by using the mass defect the formula for the relation between the energy and mass
discovered by Einstein.  We proceed in a different way, namely, we use the de Broglie wavelength to determine
whether the particle is quantum, and if this is the case, then we determine the energy at which the neutron
separates from the atomic nucleus, which results in the change of spin.
If the number of nucleons is even, then the atomic nucleus corresponds to a Bose particle.
If one nucleon is separated, then the nucleus becomes a Fermi particle with nonzero spin.

The behavior of Bose and Fermi particles is described by the Bose--Einstein and Fermi--Dirac distributions, respectively.
The Bose--Einstein distribution in polylogarithm form becomes
 \begin{equation}\label{B-E}
    \operatorname{Li}_s(a) = \frac{1}{\Gamma(s)} \int_0^\infty \frac{t^{s-1}}{e^t/a-1}\, dt,
      \end{equation}
where $\operatorname{Li}_{(\cdot)}(\cdot)$ is  the polylogarithm.
The Fermi--Dirac distribution can be written as
  \begin{equation}\label{F-D}
   - \operatorname{Li}_s(-a) = \frac{1}{\Gamma(s)} \int_0^\infty \frac{t^{s-1}}{e^t/a+1}\, dt.
      \end{equation}

We consider the quantum particles each of which is associated with a wave packet.
These wave packets are related to the de Broglie thermal wavelength~$\Lambda$.

Considering the $\Omega$-potential corresponding to the Gentile statistics~\cite{Gentile},
we can obtain a detailed description of the transition  from  the Bose gas particles of a nucleus into   the Fermi gas particles.

Let us introduce the new notation which permits determining the energy in dimensionless form.

Let $\mathfrak{v}=\Lambda^{2s}$. This quantity has the dimension of volume in the $2s$-dimensional space.
Let $\mathbf{E}=\frac{2\pi\hbar^2}{m}{V}^{-\frac{1}{s}}$. This quantity has the dimension of energy .

Now we introduce dimensionless variables, $\mathfrak{E}={E}/{\mathbf{E}}$ for the total energy
and $\mathfrak{V}={V}/{\mathfrak{v}}$ for the volume.
We note that the quantity $\mathfrak{V}^{1/D}$ is the ratio of the characteristic linear dimension
of the system ${V}^{1/D}$ to the de Broglie wavelength~$\Lambda$.

Usually, $N_i$ denotes the number of particles located at the $i$th energy level.
It is assumed that, in the case of the Fermi gas, there is at most one particle at each energy level,
and in the case of the Bose gas, the number of particles~$N_i$ at each energy level can be arbitrarily large.
We consider the Gentile statistics~\cite{Gentile} according to which, at each energy level,
the number of particles located at each energy level is bounded by the number~$k$.
In other words, the number of particles at any energy level cannot exceed the number~$k$.

The maximal number of particles at an energy level in the system
is attained for the maximal value of the activity~$a$, i.e., at the point $a=1$.
Since $\sum_{i=1}^M N_i = N$, it is obvious that $N_i\leq N$ for the Bose system.
Therefore, $k\leq N$ for the Bose system.
In the Gentile statistics, the~$k$ are integers such that $k_i<k_{i+1}$.

We assume that $k=N$ in an infinitely small neighborhood of $[N]$, where $[N]$ is the integral part of the number~$N$.

In the nonstandard analysis developed by Robinson (see~\cite{Nestandart-1}--\cite{Nestandart-2}),
the set of points infinitely close to the number $[N]$ is called the Leibnitz differential~\cite{Shepin-2}
which is understood as the length of an elementary infinitely small interval (monad).
The differential is an arbitrary infinitely small increment of a variable.

By $x$ we denote the difference $N-[N]$, i.e., $N-[N]=x<0$
($x<0$ corresponds to boson on the PZ-diagram).
We seek the expansion in a power series in~$x$ up to $O(x^2)$, which implies that $N\sim [N]$.

For the ideal gas of dimension~$D$ obeying the Gentile statistics,
i.e., in the case where, at each energy level, there can be at most~$k$ particles
($k$ is an integer), the following relation for the number of particles~$N$ is known:
\begin{equation}\label{Gent}
N=\mathfrak{V}(\operatorname{Li}_{s}(a) -\frac{1}{(k+1)^{s-1}}\operatorname{Li}_{s}(a^{k+1})) .
\end{equation}

The self-consistent relation for $x$ in a neighborhood of $[N]$ has the form
\begin{equation}\label{Np}
[N]+x=\mathfrak{V}(\operatorname{Li}_{s}(a)-\frac{1}{([N]+x+1)^{s-1}}\operatorname{Li}_{s}(a^{[N]+x+1})).
\end{equation}

The following thermodynamical formula for the energy is known:
\begin{equation}\label{Mp}
\mathfrak{E}= s \mathfrak{V}^{\frac{s+1}{s}}(\operatorname{Li}_{s+1}(a)-\frac{1}{([N]+x+1)^s}\operatorname{Li}_{s+1}(a^{[N]+x+1})).
\end{equation}

Let us perform the same transformations for $x>0$.
Then the term at the first degree of $x$ has negative sign. It corresponds to the case of Fermi system.

We note that, in the thermodynamics, $N$ is the number of molecules. In this paper, we do not consider molecules,
we only consider the nucleus, i.e., the nuclear physics. In this sense, we can say that, in our model,
the number of molecules $N$ is zero. Therefore, in contrast to the standard Gentile statistics,
we also assume that $k=0$, and we consider only the case $[N]=0$. To the numbers $N=k=0$ we apply the nonstandard analysis
and the technique of the Gentile statistics.

Using the technique of nonstandard analysis, we add a monad~$x$ to the integer~$k$.
Then expression~\eqref{Gent} is not equal to zero.

We expand the right-hand side of Eq.~\eqref{Np} in small $x\neq 0$ omitting the third-order terms:
\begin{equation}     \label{N01-0}
\begin{split}
&x=\mathfrak{V} x \bigg((s-1) \operatorname{Li}_{s}(a)-\log (a)\operatorname{Li}_{s-1}(a)\bigg)\\
&
+\mathfrak{V} \frac{1}{2} x^2 \bigg(\log ^2(a) (-\operatorname{Li}_{s-2}(a))-(s-1)
\big(s \operatorname{Li}_{s}(a)-2 \log (a)\operatorname{Li}_{s-1 }(a)\big)\bigg).
\end{split}
\end{equation}

Cancelling $x$ in both sides of~\eqref{N01-0},
we obtain an expression for~$a_0$, i.e., the value of~$a$ at which $N=0$,
for the  Bose-Einstein distribution:
\begin{equation}
\label{N=0}
(s-1) \operatorname{Li}_{s}(a_0)-  \log (a_0)\operatorname{Li}_{s-1}(a_0)-\mathfrak{V}^{-1}=0.
\end{equation}

Similarly, for the Fermi--Dirac distribution, we obtain $a_0>0$:
\begin{equation}
\label{N=0F}
(s-1)
\left(- \operatorname{Li}_{s}(-a_0)\right) -  \log (-a_0)\left(-\operatorname{Li}_{s-1}(-a_0)\right)-\mathfrak{V}^{-1}=0, \qquad a_0 <0.
\end{equation}

The value $ \operatorname{Li}_{s}(a)$,  where $a=e^{\mu/T}$, is associated with the total energy of transition,
in particular, in the three-dimensional case ($s=3/2$).

After elimination of $x$ in~\eqref{N01-0},
we consider the negative monads $x$ for the Bose--Einstein distribution
and positive monads for the Fermi--Dirac distribution, i.e., $x>0$.
As a result, when we sum the energy of decay of a Bose gas particle
and the energy obtained by the Fermi particle, then the term with $x^2$
is cancelled and we obtain the energy value~\eqref{E1}.

We note that it follows from Eq.~\eqref{N=0} that, $a_0\to 0$ as $\mathfrak{V}\to{\infty}$.
This means that the values~$a_0$ are small in the case where the value of the system characteristics linear dimension,
which is equal to~${V}^{1/D}$, exceed the de Broglie thermal wavelength~$\Lambda$.

For a sufficiently large value $\mathfrak{V} =\frac{V}{\Lambda^{2s}}$, Eq.~\eqref{N=0} has a unique solution
$a_0\le1$ which depends on $\frac{V}{\Lambda^{2s}},s$. We have
\begin{equation}
\label{N=02}
(s-1) \operatorname{Li}_{s}(a_0)-  \log (a_0)\operatorname{Li}_{s-1}(a_0)=\frac{\Lambda^{2s}}{V}.
\end{equation}

Similarly, for the Fermi--Dirac distribution.

The expression for the de Broglie thermal wavelength $\Lambda$ has the form
$\Lambda=\sqrt{\frac{2\pi\hbar^2}{m T}}$.

The value of the activity $a$ at a known temperature~$T$ determines the following value of the chemical potential~$\mu$:
\begin{equation}
\label{mu0}
\mu=T \log(a)\le0.
\end{equation}

In particular, at $a=a_0$, the greater the temperature $T$, the less $a_0$ and the greater the corresponding value $|\mu_0|$.
Thus, as the temperature increases, the transition point~$\mu_0$ approaches the point $\mu=-\infty$
at which the pressure~$P$ changes sign.

Assume that $a_0=1$ and the mass $m$ and the volume $V$ of the nucleus are known.
Then, taking the expression for the de Broglie thermal wavelength $\Lambda=\sqrt{\frac{2\pi\hbar^2}{m T}}$ into account,
we can consider Eq.~\eqref{N=02} as an equation for~$T$.

The temperature arising at $a=1$, i.e., as $\mu \to 0$, will be called the critical temperature.
We denote it by~$T_s$.
Since the temperature $T_s$ is the lowest on the whole interval of variation in~$\mu$
which is the ray $(-\infty,0]$, the ratio $T/T_s$ will be called the regularized temperature,
and we denote it by $T_{\text{reg}}$. The temperature variation along the isotherm can be measured
in~$T_{\text{reg}}$.

The expansion of the energy~\eqref{Mp} in small~$x$ up to the first order inclusively has the form
\begin{equation}
\label{E1}
\mathfrak{E}\, dx=
2s \mathfrak{V}^{\frac{1}{s}} \bigg(s \operatorname{Li}_{s+1}(a_0)-\log (a_0)\operatorname{Li}_{s}(a_0)\bigg)\, dx.
\end{equation}

The value $a_0$ is calculated by formulas~\eqref{N=0} and \eqref{N=0F}.

If we divide by $dx$, we can obtain a new formula for the specific energy.

Thus, we have calculated the energy of a neutron separation from the atomic nucleus,
i.e., the energy necessary for one neutron to leave the nucleus provided that the volume of the atomic nucleus
and the de Broglie thermal wavelength are known.

\section*{Appendix}

\begin{center}

\textbf{\large Quasi-Particles Associated with Lagrangian Manifolds\\
Corresponding to Semiclassical Self-Consistent Fields.~III}\footnote{The work
was supported by the ISF under grant No.~MFO000.}

\medskip

\textit{Victor~P.~Maslov}
\medskip

Moscow Institute of Electronics and Mathematics (Technical University),

3/12 B. Vuzovski\u\i\ per., 109028 Moscow, Russia
\medskip

Russian Journal of Mathematical Physics Vol.~3 No.~2 (1995)
\end{center}

In the preceding part of this paper, we presented Eqs.~\thetag{25} for
quasi-particles associated with an $n$-dimensional Lagrangian manifold and
Eq.~\thetag{29} for quasi-particles corresponding to a $(2n-1)$-dimensional
manifold. These equations were written out only in the $x$-chart, and the
quantum corrections were given without proof. In this part we essentially use
the canonical operator ideology to obtain Eq.~\thetag{25} with corrections in
the $x$-chart as well as in any other chart of the canonical atlas~[1].
To derive the correction in Eq.~\thetag{29}, a ``modified" $\dt$-function
must be used, and this will be done in the next part of the paper.

To obtain the result in an arbitrary canonical chart, one should pass on to
the $p$-representation with respect to some of the coordinates in the Hartree
equation. This is actually equivalent~[2] to considering the Hartree-type equation
\begin{align}
&\Big[H_0\Big(\overset 2 \to x, -\overset 1\to{ih\frac{\partial}{\partial x}}\Big)
+ \int dy \psi^*(y) H_1 \Big(\overset 2 \to x, - \overset 1\to{ih\frac{\partial}{\partial x}};
\overset 2 \to y, -\overset 1\to{ih\frac{\partial}{\partial y}}\Big)\psi(y)\Big]\psi(x)
\nonumber\\
&\qquad
= \Omega \psi(x),
\tag{50}
\end{align}
where $x,y\in\mathbb{R}^n$, $\psi\in L^2(\mathbb{R}^n)$ is a complex-valued function,
$h>0$, $\Omega\in\mathbb{R}$, and the indices $1$ and $2$ specify the ordering of the
operators $x$ and $-ih\partial/\partial x$.
The function $H_1$ satisfies the condition
$H_1(x,p_x;y,p_y)=H_1(y,p_y;x,p_x)$.
Equation~\thetag{50} generalizes the ordinary Hartree equation
(Eq.~\thetag{1} in~[4], where $N=1$).
The study of Eq.~\thetag{50} is important, for example, if one makes an
attempt to find a solution to the Hartree equation \thetag{1} in the momentum
representation,
$$
\psi(x)=\int \widetilde\psi(p)e^{(i/\hbar)px} \frac{dp}{(2\pi\hbar)^{n/2}}.
$$

Let us also discuss the variational system associated with Eq.~\thetag{50},
which can be obtained as follows.
Along with Eq.~\thetag{50}, let us write out the conjugate equation and
consider the variations of both equations {\it assuming that the variations\/}
$\dt\psi=F$ and $\dt\psi^*=G$ are  {\it independent\/}.

The variational system has the form
\begin{align}
&\Big[H_0\Big(\overset 2 \to x, -\overset 1\to{ih\frac{\partial}{\partial x}}\Big)
-\Omega + \int dy \psi^*(y) H_1
\Big(\overset 2 \to x, -\overset 1\to{ih\frac{\partial}{\partial x}};
\overset 2 \to y, -\overset 1\to{ih\frac{\partial}{\partial y}}\Big)\psi(y)\Big]F(x)
\nonumber\\
&\quad +\int dy\Big(G(y) H_1\Big(\overset 2 \to x, -\overset 1\to{ih\frac{\partial}{\partial x}};
\overset 2 \to y, -\overset 1\to{ih\frac{\partial}{\partial y}}\Big)\psi(y)
\nonumber\\
&\quad+\psi^*(y) H_1\Big(\overset 2 \to x, -\overset 1\to{ih\frac{\partial}{\partial x}};
\overset 2 \to y, -\overset 1\to{ih\frac{\partial}{\partial y}}\Big)F(y)\Big) \psi(x)
= -\beta F(x),
\nonumber\\
&\Big[H_0\Big(\overset 1 \to x, \overset 2\to{ih\frac{\partial}{\partial x}}\Big)
-\Omega + \int dy \psi(y) H_1
\Big(\overset 1 \to x, \overset 2\to{ih\frac{\partial}{\partial x}};
\overset 1 \to y, \overset 2\to{ih\frac{\partial}{\partial y}}\Big)\psi^*(y)\Big]G(x)
\nonumber\\
&\quad +\int dy\Big(
F(y) H_1\Big(\overset 1 \to x, \overset 2\to{ih\frac{\partial}{\partial x}};
\overset 1 \to y, \overset 2\to{ih\frac{\partial}{\partial y}}\Big)\psi^*(y)
\nonumber\\
&\quad +\psi(y) H_1\Big(\overset 1 \to x, \overset 2\to{ih\frac{\partial}{\partial x}};
\overset 1 \to y, \overset 2\to{ih\frac{\partial}{\partial y}}\Big)G(y)\Big) \psi^*(x)
= \beta G(x).
\tag{51}
\end{align}
Equations~\thetag{50} and \thetag{51} play an important role in the problem of
constructing  asymptotic solutions to the $N$-particle Schr\"odinger
equation as $N\to\infty$~[5--7].

For example, the spectrum of system \thetag{51} (possible values of~$\beta$)
corresponds to the spectrum of quasi-particles. Namely, the
difference between the energy of an excited state and the ground state
energy is given by the  expression
$$
\sum_k \beta_k n_k,
$$
where the numbers $n_k\in Z_+$, $k=\overline{1,\infty}$,
which are equal to zero starting from some $k$,
define the eigenfunction and the eigenvalue of the
excited state, and $\beta_k\in\mathbb{R}$ are the eigenvalues of system
\thetag{51}.

In this paper we are interested in asymptotic solutions to
Eqs.~\thetag{50} and \thetag{51} as the ``inner" $h$ tends to zero.

Asymptotic solutions to Eq.~\thetag{50} are
given~[8] by the canonical operator on a Lagrangian manifold
$\Lambda^n=\{x=X(\alpha), p=P(\alpha)\}$ invariant
with respect to the Hamiltonian system
\begin{equation}
\dot x=\frac{\partial H(x,p)}{\partial p}, \quad
\dot p=-\frac{\partial H(x,p)}{\partial x},
\tag{52}
\end{equation}
where
$$
H(x,p)=H_0(x,p) + \int d\mu_\alpha H_1(x,p; X(\alpha),P(\alpha)),
$$
$\alpha\in\Lambda^n$, and $d\mu_\alpha$ is an invariant measure on $\Lambda^n$.
The Lagrangian manifold lies on the surface $H(x,p)=\Omega$.
If a chart $A$ is projected diffeomorphically in the $x$-plane,
then the canonical operator acts as the multiplication by
$\exp\{(i/h)S(x)\}/\sqrt J$, where $S(x)=\int p\,dx$ on
$\Lambda^n$ and $J=Dx/D\mu_\alpha$. We are interested in finding asymptotic
solutions to Eqs.~\thetag{51}. Without loss of generality, we can consider only
the case of $x$-chart. Indeed, to obtain similar expressions in the
$p$-chart, one must consider the Fourier transformation of Eqs.~\thetag{50} and
\thetag{51} and apply the same technique, since the form of the equations
remains unchanged.

Let us seek the asymptotic solutions to Eqs. \thetag{51} in the
$x$-chart in the form
\begin{equation}
F(x) = \widetilde f(x)\psi(x),\quad G(x)=\widetilde g(x)\psi^*(x),
\tag{53}
\end{equation}
where the functions $f$ and $g$, in contrast to $\psi$ and $\psi^*$, have a
limit as $h\to0$. One can consider a more general case, by allowing $f$ and $g$
to be functions of $x$ and $-ih\partial/\partial x$, but in the leading term as $h\to0$ we
have
$$
-ih\frac{\partial}{\partial x} e^{(i/h)S}\approx \frac{\partial S}{\partial x}e^{(i/h)S},
$$
and so we arrive at functions $f$ and $g$ that depend only on $x$.

The second equation in system \thetag{51} can be rewritten in the form
\begin{align}
&\Big[H_0\Big(x, {ih\frac{\partial}{\partial x}} \Big)
+ \int dy \psi(y) H_1 \Big( x,{ih\frac{\partial}{\partial x}};
y, {ih\frac{\partial}{\partial y}}\Big)\psi^*(y);\widetilde g(x)\Big]\psi^*(x)
\nonumber\\
&\quad +\int dy\Big\{ \psi(y)\widetilde c(y) H_1\Big(x, ih\frac{\partial}{\partial x}; y,
ih\frac{\partial}{\partial y}\Big)\psi^*(y)
\nonumber\\
&\quad+\psi(y) \Big[H_1\Big( x, {ih\frac{\partial}{\partial x}};
y, {ih\frac{\partial}{\partial y}}\Big);\widetilde g(y)\Big] \psi^*(y)\Big\}
= \beta\widetilde g(x) \psi^*(x),
\tag{54}
\end{align}
where $[A;B]=AB-BA$ and
\begin{equation}
c(x)=\widetilde f(x)+\widetilde g(x).
\tag{55}
\end{equation}
Equation~\thetag{50} is used in the derivation of Eq.~\thetag{54}. We observe that all terms
containing the function $\widetilde g$ on the left-hand side in Eq.~\thetag{50} are $O(h)$, since the
commutator of two operators depending on $x$ and $-ih\partial/\partial x$ is  equal in the classical
limit to $(-ih)$ times the Poisson bracket of the corresponding classical quantities.

Thus, the function $c$, as well as the eigenvalue $\beta$, is assumed to be
$O(h)$. Let us rescale these quantities as follows:
\begin{equation}
c(x)=h\widetilde c(x),\quad\beta=h\widetilde\beta.
\tag{56}
\end{equation}

Now we can derive the equation for $\widetilde g, \widetilde c$, and $\widetilde\beta$
in the leading term in $h$ and the first correction to it from Eq.~\thetag{54},
making use of the following relations:
\begin{align}
\text{i)}\,\,\, \Big[A\Big(\overset 1 \to x, \overset 2\to{ih\frac{\partial}{\partial x}}\Big);
\xi(x)\Big]
&=\sum_{a=1}^n i h \frac{\partial A}{\partial p_a} \Big(\overset 1 \to x,
\overset 2\to{ih\frac{\partial}{\partial x}}\Big)\frac{\partial\xi}{\partial x_a}
\nonumber\\
&\qquad
- \sum_{a,b=1}^n \frac{h^2}{2}\frac{\partial^2A}{\partial p_a \partial p_b}
\Big(\overset 1 \to x, \overset 2\to{ih\frac{\partial}{\partial x}}\Big)
\frac{\partial^2\xi}{\partial x_a \partial x_b},
\tag{57}
\end{align}
where $p_a=ih\partial/\partial x_a$, $A(x,p)$ is a function $\mathbb{R}^{2n}\to\mathbb{C}$,
$\xi:\mathbb{R}^n\to\mathbb{R}$.

{\hskip 0.3 cm}ii) $\psi(x)=\chi(x,h)e^{(i/h)S(x)}$, where $\chi=1/\sqrt J$ in
the leading term in $h$;
$$
\text{iii)}\qquad\qquad\qquad\qquad
ih\frac{\partial}{\partial x}e^{-(i/h)S(x)} = e^{-(i/h)S(x)}
\Big(\frac{\partial S}{\partial x_n}+ih \frac{\partial}{\partial x}\Big);\qquad\qquad\qquad\quad
\hfill
$$
\begin{align}
\text{iv)}\quad\quad\,\,
&B\Big(ih\frac{\partial}{\partial x} + \frac{\partial S}{\partial
x}\Big) = B \Big(\frac{\partial S}{\partial x}\Big) + ih \sum_{a=1}^n \frac{\partial B}{\partial
p_a} \frac{\partial}{\partial x_a} + \frac{ih}{2}\sum_{a,b=1}^n \frac{\partial^2 B}{\partial
p_a\partial p_b} \frac{\partial^2 S}{\partial x_a\partial x_b}
\nonumber\\
&\quad+\frac{(ih)^2}{2}\sum_{a,b=1}^n \frac{\partial^2 B}{\partial p_a\partial p_b}
\frac{\partial^2}{\partial x_a\partial x_b} + \frac{(ih)^2}{2}\sum_{a,b,c=1}^n \frac{\partial^3
B}{\partial p_a\partial
p_b \partial p_c} \frac{\partial^2 S}{\partial x_a\partial x_b}\frac{\partial}{\partial x_c}
\nonumber\\
&\quad+\frac{(ih)^2}{6}\sum_{a,b,c=1}^n \frac{\partial^3 B}{\partial p_a\partial
p_b \partial p_c} \frac{\partial^3 S}{\partial x_a\partial x_b\partial x_c}
\nonumber\\
&\quad +\frac{(ih)^2}{8}\sum_{a,b,c,d=1}^n \frac{\partial^4 B}{\partial p_a\partial p_b \partial p_c
\partial p_d} \frac{\partial^2 S}{\partial x_a\partial x_b}
\frac{\partial^2 S}{\partial x_c\partial x_d} +O(h^3),
\tag{58}
\end{align}
where all derivatives of $B$ are evaluated at the point $p=\partial S/\partial x$.

These relations can  easily be obtained for monomial functions $A$ and
$B$. An application of formulas i)--iv) yields the  equation
\begin{align}
&i\sum_{a=1}^n \frac{\partial H}{\partial p_a^x}\frac{\partial \widetilde g}{\partial x_a}(X(\alpha))
- \widetilde\beta\widetilde g (X(\alpha)) + \int d\mu_\beta \widetilde c(X(\beta))H_1
\nonumber\\
&\quad + i\int d\mu_\beta \sum_{a=1}^n \frac{\partial \widetilde g}{\partial x_a}(X(\beta))
\frac{\partial H_1}{\partial p_a^y}
 +\frac{h}{2} \sum_{a,b=1}^n \frac{\partial \widetilde g}{\partial x_a}(X(\alpha))
\frac{\partial^2 H}{\partial p_a^x \partial p_b^x}\frac{\partial\ln J}{\partial x_b}(X(\alpha))
\nonumber\\
&\quad -\frac{h}{2} \sum_{a,b,c=1}^n \frac{\partial \widetilde g}{\partial x_a}(X(\alpha))
\frac{\partial^3 H}{\partial p_a^x \partial p_b^x \partial p_c^x}\frac{\partial^2S}{\partial x_b
\partial x_c}(X(\alpha))
 -\frac{h}{2} \sum_{a,b=1}^n \frac{\partial^2 \widetilde g}{\partial x_a \partial
x_b}(X(\alpha)) \frac{\partial^2 H}{\partial p_a^x \partial p_b^x}
\nonumber\\
&\quad + \frac{ih}{2} \int d\mu_\beta \widetilde c(X(\beta)) \sum_{a,b=1}^n \Big[\frac{\partial^2
H_1}{\partial p_a^x \partial p_b^x} \frac{\partial^2 S}{\partial x_a \partial x_b}(X(\alpha)) +
\frac{\partial^2 H_1}{\partial p_a^y \partial p_b^y}
\frac{\partial^2 S}{\partial x_a \partial x_b}(X(\beta))\Big]
\nonumber\\
&\quad - \frac{ih}{2} \int d\mu_\beta \widetilde c(X(\beta)) \sum_{a=1}^n \Big[\frac{\partial
H_1}{\partial p_a^x} \frac{\partial\ln J}{\partial x_a}(X(\alpha))
+ \frac{\partial H_1}{\partial p_a^y}
\frac{\partial\ln J}{\partial x_a}(X(\beta))\Big]
\nonumber\\
&\quad + \frac{h}{2} \int d\mu_\beta  \sum_{a=1}^n \frac{\partial\widetilde g}{\partial x_a}(X(\beta))
\Big\{\sum_{b=1}^n\Big(\frac{\partial^2 H_1}{\partial p_a^y \partial p_b^y}
\frac{\partial\ln J}{\partial x_b}(X(\beta)) + \frac{\partial^2 H_1}{\partial p_a^y \partial p_b^x}
\frac{\partial\ln J}{\partial x_b}(X(\alpha))\Big)
\nonumber\\
&\quad - \sum_{b,c=1}^n \Big(\frac{\partial^3 H_1}{\partial p_a^y \partial p_b^x \partial p_c^x}
\frac{\partial^2 S}{\partial x_b \partial x_c}(X(\alpha)) + \frac{\partial^3 H_1}{\partial p_a^y
\partial p_b^y \partial p_c^y}
\frac{\partial^2 S}{\partial y_b \partial y_c}(X(\beta))\Big\}
\nonumber\\
&\quad -\frac{h}{2} \int d\mu_\beta \sum_{a,b=1}^n \frac{\partial^2 H_1}{\partial p_a^y \partial
p_b^y}\frac{\partial^2 g}{\partial x_a \partial x_b}(X(\beta)) =0;
\tag{59}
\end{align}
in this formula the arguments
\begin{equation}
x=X(\alpha),\quad p^x=P(\alpha),\quad y=X(\beta),\quad p^y=P(\beta)
\tag{60}
\end{equation}
of the function $H_1$ and of its derivatives, as well as the arguments
$x=X(\alpha)$, $p^x=P(\alpha)$ of the function $H$, are omitted.

Let us now find another equation relating $\widetilde g$ to $\widetilde c$. To this end, let us
multiply the first equation in system \thetag{51} by $\psi^*(x)$ and the second equation by
$\psi(x)$. Let us subtract  the first product from the second. We obtain
\begin{align}
&\beta\psi^*(x) \psi(x)c(x) = \psi(x)
\Big[H\Big(x,ih\frac{\partial}{\partial x}\Big); \widetilde g(x)\Big]\psi^*(x)
\nonumber\\
&\quad+  \psi^*(x) \Big[H\Big(x,-ih\frac{\partial}{\partial x}\Big); \widetilde g(x)\Big]\psi(x)
- \psi^*(x)
\Big[H\Big(x,-ih\frac{\partial}{\partial x}\Big); \widetilde c(x)\Big]\psi(x)
\nonumber\\
&\quad + \psi(x)\int dy \psi(y) \Big[H_1\Big(x,ih\frac{\partial}{\partial x}; y,
ih\frac{\partial}{\partial y}\Big); \widetilde
g(y)\Big]\psi^*(y)\psi^*(x)
\nonumber\\
&\quad - \psi^*(x)\int dy \psi^*(y) \Big[\widetilde g(y); H_1\Big(x,-ih\frac{\partial}{\partial x};
y, -ih\frac{\partial}{\partial y}\Big)\Big]\psi(y)\psi(x)
\nonumber\\
&\quad + \psi(x)\int dy \psi(y) c(y)
H_1\Big(x,ih\frac{\partial}{\partial x}; y, ih\frac{\partial}{\partial y}\Big)\psi^*(y)\psi^*(x)
\nonumber\\
&\quad - \psi^*(x)\int dy \psi^*(y) H_1\Big(x,-ih\frac{\partial}{\partial x}; y,
-ih\frac{\partial}{\partial y}\Big) c(y)\psi(y)\psi(x).
\tag{61}
\end{align}

Let us use Eqs. \thetag{57}--\thetag{59}.
We find the following equation for $\widetilde g$ and
$\widetilde c$ modulo $O(h^2)$:
\begin{align}
&i\sum_{a=1}^n \frac{\partial H}{\partial p_a^x}\frac{\partial \widetilde c}{\partial
x_a}(X(\alpha)) - \widetilde\beta\widetilde c (X(\alpha)) - \sum_{a,b=1}^n \frac{\partial^2 H}{\partial p_a^x
\partial p_b^x} \frac{\partial^2 \widetilde g}{\partial x_a \partial x_b}(X(\alpha))
\nonumber\\
&\quad + \sum_{a,b=1}^n \frac{\partial^2 H}{\partial p_a^x \partial p_b^x}\frac{\partial \widetilde
g}{\partial x_a}(X(\alpha)) \frac{\partial\ln J}{\partial x_b}(X(\alpha))
\nonumber\\
&\quad - \sum_{a,b,c=1}^n \frac{\partial \widetilde g}{\partial x_a}(X(\alpha)) \frac{\partial^3
H}{\partial p_a^x \partial p_b^x \partial p_c^x}\frac{\partial^2S}{\partial x_b \partial x_c}(X(\alpha))
- \int d\mu_\beta \sum_{a,b=1}^n \frac{\partial^2 \widetilde g}{\partial
x_a \partial x_b}(X(\beta)) \frac{\partial^2 H_1}{\partial p_a^y \partial p_b^y}
\nonumber\\
&\quad - \int d\mu_\beta \sum_{a,b,c=1}^n \frac{\partial\widetilde g}{\partial x_a}(X(\beta))
\Big(\frac{\partial^2 S}{\partial x_a \partial x_b}(X(\alpha)) \frac{\partial^3 H_1}{\partial p_a^y
\partial p_b^x \partial p_c^x} + \frac{\partial^2 S}{\partial y_a \partial y_b}(X(\beta))
\frac{\partial^3 H_1}{\partial p_a^y \partial p_b^y \partial p_c^y}\Big)
\nonumber\\
&\quad +\int d\mu_\beta \sum_{a,b=1}^n \frac{\partial \widetilde g}{\partial x_a}(X(\beta))
\Big(\frac{\partial\ln J}{\partial x_b}(X(\alpha)) - \frac{\partial^2 H_1}{\partial p_a^y \partial
p_b^x} + \frac{\partial\ln J}{\partial x_b}(X(\beta))\frac{\partial^2 H_1}{\partial p_a^y \partial
p_b^y}\Big)
\nonumber\\
&\quad - i\int d\mu_\beta \widetilde c(X(\beta)) \sum_{a=1}^n \Big(\frac{\partial\ln J}{\partial
x_a}(X(\alpha))\frac{\partial H_1}{\partial p_a^x} + \frac{\partial\ln J}{\partial x_a}(X(\beta))
\frac{\partial H_1}{\partial p_a^y}\Big)
\nonumber\\
&\quad+ i \int d\mu_\beta \sum_{a=1}^n\frac{\partial\widetilde c}{\partial x_a}(X(\beta))
\frac{\partial H_1}{\partial p_a^y}
\nonumber\\
&\quad+ i\int d\mu_\beta \widetilde c(X(\beta))\sum_{a,b=1}^n \Big(\frac{\partial^2 H_1}{\partial p_a^x
\partial p_b^x}\frac{\partial^2S}{\partial x_a \partial x_b}(X(\alpha))
+ \frac{\partial^2 H_1}{\partial p_a^y \partial
p_b^y}\frac{\partial^2S}{\partial y_a \partial x_b}(X(\beta))\Big)
\nonumber\\
&\quad + \frac{h}{2} \sum_{a,b,c=1}^n \frac{\partial\widetilde c}{\partial x_a}(X(\alpha))
\frac{\partial^3H}{\partial p_a^x \partial p_b^x \partial p_c^x}\frac{\partial^2S}{\partial x_b
\partial x_c} (X(\alpha))
\nonumber\\
&\quad - \frac{h}{2} \sum_{a,b=1}^n \frac{\partial\widetilde c}{\partial x_a}(X(\alpha))
\frac{\partial^2 H}{\partial p_a^x \partial p_b^x} \frac{\partial\ln J}{\partial x_b}(X(\alpha))
+ \frac{h}{2}\sum_{a,b=1}^n \frac{\partial^2 H}{\partial p_a^x \partial
p_b^x}\frac{\partial^2\widetilde c}{\partial x_a \partial x_b}(X(\alpha))
\nonumber\\
&\quad + \frac{h}{2} \int d\mu_\beta \sum_{a,b,c=1}^n \frac{\partial\widetilde c}{\partial
x_a}(X(\beta)) \Big( \frac{\partial^3H_1}{\partial p_a^y \partial p_b^x \partial
p_c^x}\frac{\partial^2S}{\partial x_b \partial x_c} (X(\alpha))
+ \frac{\partial^3H_1}{\partial p_a^y \partial p_b^y \partial p_c^y}
\frac{\partial^2S}{\partial x_b \partial x_c} (X(\beta))\Big)
\nonumber\\
&\quad +  \frac{h}{2} \int d\mu_\beta \sum_{a,b=1}^n \Big[\frac{\partial^2 H_1}{\partial p_a^y
\partial p_b^y}\Big( \frac{\partial^2\widetilde c}{\partial x_a \partial x_b}(X(\beta)) -
\frac{\partial\widetilde c}{\partial x_a}(X(\beta)) \frac{\partial\ln J}{\partial x_b}(X(\beta))
\nonumber\\
&\quad - \frac{\partial\widetilde c}{\partial x_a}(X(\beta)) \frac{\partial\ln J}{\partial
x_b}(X(\alpha)) \frac{\partial^2 H_1}{\partial p_a^y \partial p_b^x}\Big]=0.
\tag{62}
\end{align}
If
$$
H_0(x,p_x) = p_x^2/2 +U(x),\qquad
H_1(x,p_x;y,p_y)= V(x,y),
$$
then Eqs.~\thetag{59} and \thetag{62} become much  simpler and acquire the form
\begin{align}
(i\nabla S\nabla -\widetilde\beta)\widetilde g
+ \int V(x,X(\alpha'))\widetilde c(X(\alpha'))\,d\mu_{\alpha'}
+ \frac{h}{2}(-\Delta\widetilde g+\nabla\ln J\nabla \widetilde g)&=0,
\nonumber\\
(i\nabla S\nabla -\widetilde\beta)\widetilde c -\Delta\widetilde g +\nabla \ln J\nabla \widetilde g
- \frac{h}{2}(-\Delta\widetilde c+\nabla\ln J\nabla \widetilde c)&=0.
\tag{63}
\end{align}
From Eqs.~\thetag{63} one can approximately find the functions $F$ and $G$,
which are important for constructing  approximate  wave functions in the
$N$-particle problem as $N\to\infty$~[5].

Let us now relate the obtained results to the solution to variational equation
for the Vlasov equation, obtained in the preceding part of this paper~[3].

Let $\widehat\rho$ be the projection on the function $\psi$. Its kernel is
$\widetilde\rho(x,y)=\psi(x)\psi^*(y)$, and its symbol is $\rho(x,p) =\psi(x)\widetilde\psi^*(p)
e^{(i/\hbar)px}$.
The operator $\widehat\rho$ satisfies the Wigner equation, which  reduces to the Vlasov
equation as $h\to0$. The operator $\widehat\sigma$ with the kernel $F(x)\psi^*(y)+\psi(x)G(y)$
is equal to
\begin{equation}
\widehat\sigma = \widetilde f\widehat\rho +\widehat\rho\widetilde g
\tag{64}
\end{equation}
and satisfies the variational equation to the Wigner equation, which is reduced to the variational
equation for the Vlasov equation \thetag{20}. In Eq.~\thetag{64} $\widetilde f$ and $\widetilde g$
are the operators of multiplication by the functions $\widetilde f$ and $\widetilde g$. We see that
in the semiclassical approximation the symbol of $\sigma$ is $O(h)$, since
$$
\widehat\sigma=[\widehat\rho;\widetilde g]+\hbar\,\widetilde c\, \widehat\rho
$$
and
$$
\sigma(x,p) \simeq \hbar\Big(-i\sum_{a=1}^n \frac{\partial\rho}{\partial p_a}(x,p)
\frac{\partial\widetilde g}{\partial x_a}+\widetilde c\rho\Big).
$$
Since $\rho$ is the $\dt_\Lambda$-function in the semiclassical
approximation~[3], the function $\sigma$ is actually the sum of the
$\dt_\Lambda$-function and its derivative. Equations~\thetag{63} are consistent
with Eqs.~\thetag{24}, obtained in~[3]
for the coefficients of $\dt$ and $\dt'$. Thus, the approach suggested in this part
of the article allows us to find an asymptotic formula for $\sigma$ as well.

The author is deeply grateful to O.~Yu.~Shvedov, whose assistance in carrying
out all computations was invaluable.

\begin{center}
\textbf{References}
\end{center}

[1] Maslov. V. P.
\textit{Th\'eorie des Perturbations et M\'ethodes Asymptotiques}
(Dunod, Paris, 1972).

[2] Maslov. V. P.
``Equations of self-consistent field'',
in \textit{Sovremennye problemy matematiki}
(1978), Vol.~11, pp.~153--234.

[3] Maslov. V. P.
``Quasi-particles associated with Lagrangian manifolds corresponding
to classical self-consistent fields, II,''
Russian J. of Math. Phys. \textbf{3} (1), 123--132 (1995).

[4] Maslov. V. P.
``Quasi-particles associated with Lagrangian manifolds correspondijng
to classical self-con\-sistent fields, I,''
Russian J. of Math. Phys. \textbf{2} (4), 528--534 (1995).

[5] Maslov. V. P. and Shvedov, O. Yu.
``Quantization in the neighborhood of classical solution in the
$N$-particle problem and superfluidity,''
Theoret. and  Math. Phys. \textbf{98} (2), 181--196 (1994).

[6] Maslov. V. P. and Shvedov, O. Yu.
``Complex WKB-method in the Fock space,''
Dokl. Ross. Akad. Nauk \textbf{340} (1), 42--47 (1995).

[7] Maslov. V. P. and Shvedov, O. Yu.
``Large deviations in the many-body problem,''
Matem. Zametki \textbf{57} (1), 133--137 (1995).

[8] Maslov. V. P.
\textit{Complex Markov Chains and Feynman Path Integral}
(Nauka, Moscow, 1976).


\begin{thebibliography}{99}

\bibitem{Landau_St-ph}
 L.~D.~Landau
and E.~M.~Lifshits, \textit{Statistical Physics} (Nauka, Moscow, 1964) [in
Russian].

\bibitem{Landau_Quan-mech}
 L.~D.~Landau and E.~M.~Lifshits,
\textit{Quantum Mechanics} (Nauka,
 Moscow,
1976) [in Russian].

  \bibitem{Gentile}
W.-S. Dai, M.Xie,  ``Gentile
 statistics with a large maximum  occupation number,''
Annals of Physics \textbf{309}, 295--305 (2004).

\bibitem{Kvasn}
I.~A.~Kvasnikov, {\it Thermodynamics and Statistical Physics: Theory of
Equilibrium Systems} (URSS, Moscow, 2002), Vol.~2 [in Russian].

\bibitem{FAN-2003}
V.~P.~Maslov,   ``Mathematical Aspects of Weakly Nonideal Bose and Fermi Gases
on a Crystal Base'',
 Funktsional.
 Anal. i Prilozhen.
\textbf{37} (2), 16--27 (2003) [Functional Anal.
 Appl. \textbf{37}
(2), (2003)].

\bibitem{Erdos-Leh}
P.~Erd{\H o}s, J.~Lehner, ``The Distribution of the  Number of Summands in the
Partitions of a Positive Integer,''
 Duke Math. J.
\textbf{8} (2), 335--345 (June 1941).

\bibitem{RJ_19-1}
 V.~P.~Maslov,
``New Probability Theory Compatible with the New Conception of Modern
Thermodynamics: Economics and Crisis of Debts,'' Russian Journal of Math.
Physics \textbf{19} (1), 63--100 (2012).

\bibitem{Erdos-46}
 P.~Erd\H{o}s, ``On some asymptotic formulas in the theory of partitions,''
 Bull. Amer. Math. Soc.
\textbf{52}, 185--188, (1946).


\bibitem{MN_91-5}
V.P.Maslov, ``Unbounded Probability Theory Compatible with the Probability
Theory of Numbers,'' Math. Notes,  \textbf{91} (5)  603--609,   (2012).


\bibitem{MN-85-1}
V.~P.~Maslov, ``Theorems on the  Debt Crisis and the Occurrence of Inflation,''
Math. Notes,  \textbf{85} (1) 146--150, (2009).


\bibitem{Masl_Shvedov}
V.~P.~Maslov and O.~Yu.~Shvedov, \textit{The Method of Complex Germ}. (URSS,
Moscow, 2000)[in Russian].


\bibitem{VKB_Metod}
 V.~P.~Maslov,
\textit{The Complex WKB Method in Nonlinear Equations} (Nauka,  Moscow, 1977)
[in Russian];
[V.~P.~Maslov, \textit{The Complex WKB Method for Nonlinear Equations I}
(Birkh\"auser Verlag, Basel--Boston--Berlin, 1994)].

\bibitem{MN_55-3}
V.~P.~Maslov,
``On an Integral Equation of the Form
 $u(x)= F(x)+\int G(x,\xi)  u^{(n-2)/2}_+(\xi)d\xi / \int u^{(n-2)/2}_+(\xi)d\xi$
for $n=2$ and $n=3$'', Mat. Zametki \textbf{55} (3), 96--108 (1994) [Math.
Notes \textbf{55} (3--4), 302--311 (1994)].

\bibitem{MN_58-6}
V.~P.~Maslov, ``Spectral Series, Superfluidity, and High-Temperature
Superconductivity'', \textbf{58} (6), 933--936 (1995) [Math. Notes \textbf{58}
(5--6),  1349--1352 (1995)].

\bibitem{Masl-Shved-1997}
V.~P.~Maslov and O.~Yu.~Shvedov,
``The number of Bose-condensed particles in a weakly nonideal Bose gas,'' Mat.
Zametki \textbf{61} (5), 790--792 (1997) [Math. Notes, \textbf{61} (5),
661--664 (1997)].

\bibitem{FAN-1999}
V.~P.~Maslov,
``On an averaging method for the quantum many-body problem,'' Funktsional.
Anal. i Prilozhen. \textbf{33} (4), 50--64  (1999) [Funct. Anal. Appl.
\textbf{33} (4), 280--291 (2000)].

\bibitem{MN_93-1:v570}
 V.~P.~Maslov,
``The Mathematical Theory of Classical Thermodynamics,''
 Math. Notes \textbf{93} (1), 102--136 (2013).

\bibitem{MN_93-3:v570}
 V.~P.~Maslov,
``The unbounded theory of probability
and multistep relaxation processes,''
 Math. Notes \textbf{93} (3) 451--459 (2013).

\bibitem{Arxiv_2009:v570}
 V.~P.~Maslov,
\textit{Threshold Levels
in
Economics},
\texttt{\tt \texttt{arXiv:}0903.4783v2 [q-fin.
 ST], 3~Apr 2009}.


 \bibitem{Nestandart-1}
A. Robinson,  \emph{Non-standard analysis}
(North-Holland Publishing Co., Amsterdam, 1966).

\bibitem{Nestandart-2}
V.V.Kanovei,  M. Reeken,
 \emph{Nonstandard Analysis, Axiomatically}
 (Springer, 2004).


\bibitem{Shepin-2}
E. V.Shchepin,  ``The Leibniz differential and the Perron?Stieltjes integral,''
J. Math. Sci. \textbf{233} (1), 157--171  (2018).


\end{thebibliography}
\end{document}